%% file: neuromuscular_arxiv.tex
\documentclass{ifacconf}

\usepackage{graphicx}      % include this line if your document contains figures
\usepackage{natbib}        % required for bibliography
\usepackage{amsfonts, dsfont, mathtools, amsmath, amssymb, bbold, mathrsfs, bm, upgreek}
\usepackage[usenames]{color}
\usepackage{times} % assumes new font selection scheme installed
\usepackage{tabularx}
\usepackage{epstopdf}
\usepackage{subfigure}
\usepackage{float}
\input{definitions}

\begin{document}
\begin{frontmatter}

\title{Modeling the Neuromuscular Control System of an Octopus Arm\thanksref{footnoteinfo}} 

\thanks[footnoteinfo]{We gratefully acknowledge financial support from ONR MURI N00014-19-1-2373, and NSF EFRI C3 SoRo $\#$1830881. We also acknowledge computing resources provided by the Extreme Science and Engineering Discovery Environment (XSEDE), which is supported by National Science Foundation grant number ACI-1548562, through allocation TG-MCB190004.}

\author[mechse,csl]{Tixian Wang} 
\author[csl]{Udit Halder} 
\author[neuro]{Ekaterina Gribkova}
\author[ncsa,igb]{Mattia Gazzola}
\author[mechse,csl]{Prashant G. Mehta}

\address[mechse]{Department of Mechanical Science and Engineering}
%University of Illinois at Urbana-Champaign, 
%   Champaign, IL 61820 USA (e-mail: tixianw2@illinois.edu)}
\address[csl]{Coordinated Science Laboratory}
\address[neuro]{Neuroscience Program} 
\address[ncsa]{National Center for Supercomputing Applications}
\address[igb]{Carl R. Woese Institute for Genomic Biology \\ 
   University of Illinois at Urbana-Champaign, Champaign, IL 61820 USA \\
	Corresponding e-mail: \texttt{udit@illinois.edu}}
	 %(e-mail: author@lamar. colostate.edu)}
%\address[Third]{University of Illinois at Urbana-Champaign, 
%	Champaign, IL 61820 USA (e-mail: author@snu.ac.kr)}
%\address[Fourth]{University of Illinois at Urbana-Champaign, 
%	Champaign, IL 61820 USA (e-mail: author@snu.ac.kr)}

\begin{abstract}                % Abstract of not more than 250 words.
The octopus arm is a neuromechanical system that involves a complex interplay between peripheral nervous system (PNS) and arm musculature. This makes the arm capable of carrying out rich maneuvers. In this paper, we build a model for the PNS and integrate it with a muscular soft octopus arm. The proposed neuromuscular architecture is used to qualitatively reproduce several biophysical observations in real octopuses, including curled rest shapes and target-directed arm reaching motions. Two control laws are proposed for target-oriented arm motions, and their performance is compared against a benchmark. Several analytical results, including rest-state characterization and stability properties of the proposed control laws, are provided.
\end{abstract}


\end{frontmatter}
%===============================================================================

\section{Introduction}
The octopus arm is a muscular hydrostatic limb characterized by virtually infinite degrees of freedom and extreme flexibility~\citep{kennedy2020octopus,levy2017motor}. Octopus arm movements have been widely studied by biologists over the past few decades~\citep{gutfreund1998patterns,sumbre2001control,yekutieli2005dynamic,hanassy2015stereotypical}. Efficient control of such  flexible appendages is challenging due to the distributed nature of sensing and motor control. Several control strategies have been proposed for octopus arm movements, including stiffening wave actuation~\citep{yoram2002move,yekutieli2005dynamic,yekutieli2005dynamic2,wang2022control}, energy shaping control~\citep{chang2020energy,chang2021controlling, chang2022energy}, optimal control~\citep{cacace2019control, wang2021optimal}, and sensory feedback control~\citep{wang2022sensory}. However, these studies remain at muscle actuation level, lacking integration with the peripheral nervous system (PNS).

Beyond octopus arms, neural control has been studied over decades for various animals and applications in robotics, such as neural modeling from biological studies~\citep{ekeberg1993combined,matsuoka1984dynamic}, robot and actuators~\citep{ijspeert2007swimming,aydin2019neuromuscular}, neuromorphic control~\citep{folgheraiter2019neuromorphic,polykretis2022bioinspired}, central pattern generator for locomotion~\citep{sfakiotakis2007neuromuscular,liu2008central,wang2020bio}, etc. 

The main contribution of this paper lies in the derivation of a simple PNS model and in its assimilation within the arm musculature. The coupled neuromuscular system is then shown to recapitulate a range of experimentally observed behaviors.

%is to build a model for the octopus arm PNS and its assimilation with the arm musculature. The combined neuromuscular system is then shown to explain several experimental behavioral observations. 

\subsection{Biological background}
\begin{figure*}[t]
	\centering
	\includegraphics[width=\textwidth, trim = {0pt 0pt 0pt 0pt}]{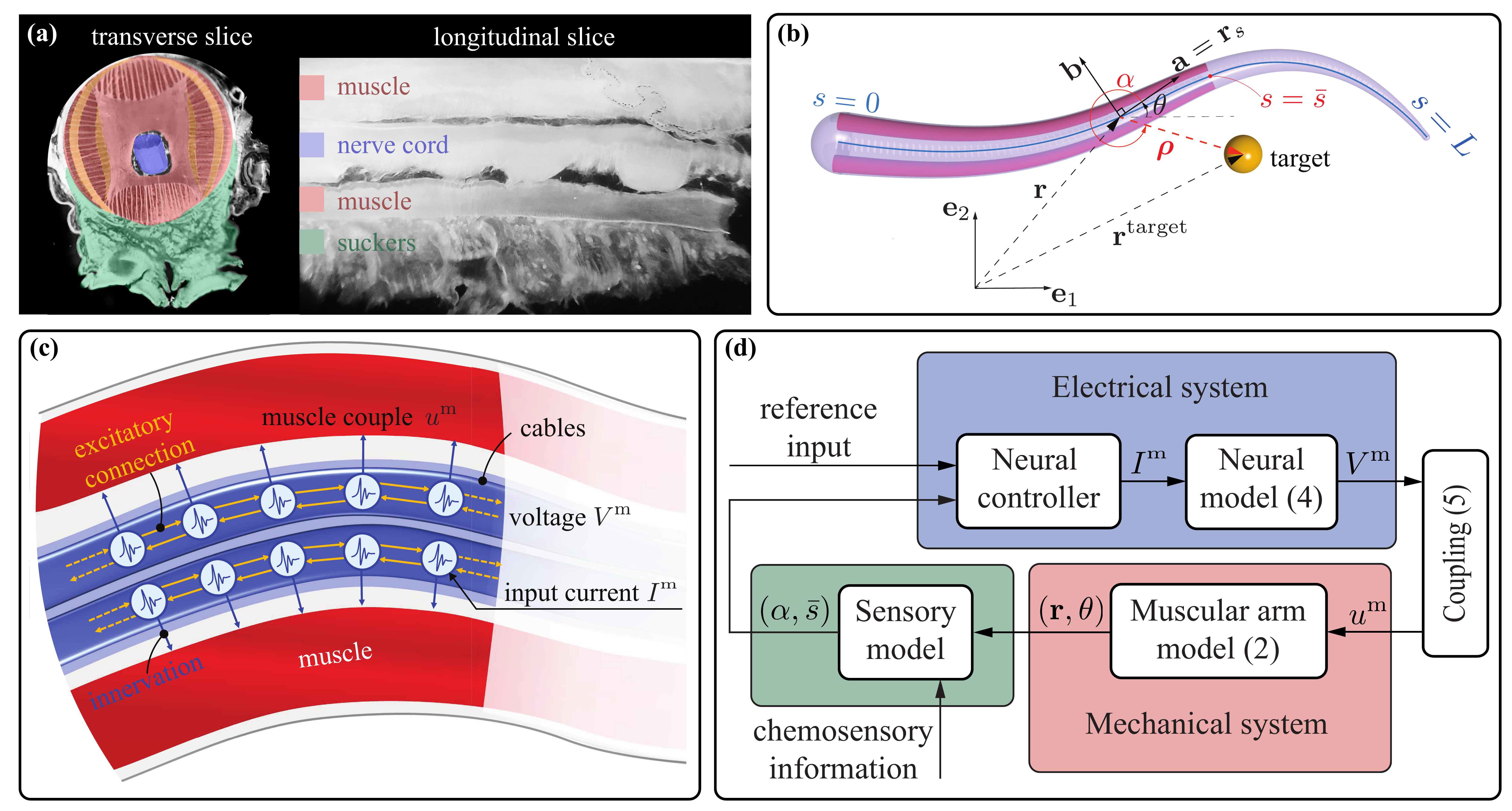}
	\caption{(a) Physiological cross sections of an \textit{Octopus rubescens} arm. The muscles are color coded in red, the nerve cord is color coded in blue, and the suckers are color coded in green. (b) Schematic of a planar octopus arm with a static target. %The top and bottom longitudinal muscles $\LMt$ and $\LMb$ are depicted in red and the axial nerve cord (ANC) is in blue. 
	The sensory information are the bearing $\alpha$ and the arc-length $\bar{s}$ of the closest point to the target. (c) Neuromuscular control architecture. The axial nerve cord is modeled by two electrical cables. (d) A block diagram for the neuromuscular control system with sensing.
	%\udit{Fig. 1 tasks: separate a), b), c) by blocks, change the font types, make c) nicer.}
	}
	\label{fig:neuromuscular_model}
	\vspace{0pt}
\end{figure*}

\noindent
{\bf 1) Anatomy:}  \textit{Neural architecture.}
The arm PNS includes a large axial nerve cord (ANC) running along the length of the arm (Fig.~\ref{fig:neuromuscular_model}(a)), and with a brachial ganglion and sucker ganglion for each sucker of the arm~\citep{rowell1963excitatory,matzner2000neuromuscular}. %associated with each sucker of the arm 
%as shown in Fig.~\ref{fig:neuromuscular_model}(a). 
It is estimated that \textit{Octopus vulgaris} has around 300 suckers arranged in two rows along each arm, with about 350 million neurons in the PNS of all eight arms~\citep{young1971anatomy, wells1978octopus}. %Spacing of suckers and their associated ganglia along the arm closely correlates with sucker diameters.
The sucker ganglia receive chemosensory information~\citep{graziadei1976sensory, mather2021octopus}, whereas the brachial ganglia receive local proprioceptive information~\citep{grasso2014octopus} and send motor commands along numerous nerve roots to the musculature of the arm~\citep{matzner2000neuromuscular}. 
Each muscle cell is intricately innervated by motorneurons for efficient and independent motor control %over muscle stiffness and contraction/actuation dynamics~
\citep{matzner2000neuromuscular,nesher2019synaptic}. 
%Furthermore, there is bidirectional communication, consisting of both sensory and motor feedback, between the sucker ganglia and brachial ganglia of the axial nerve cord. 
%The axial nerve cord also contains cerebrobrachial nerves which send sensory information to the octopus’s central nervous system (CNS) and carry descending signals from CNS to brachial ganglia. %This may enable the CNS to initiate and modulate motor programs in the arm, much like the descending control in spinal cords, such as that of the lamprey. %Many of these neuromuscular relations were elucidated through lesion and electrophysiological studies of isolated octopus arm preparations\citep{rowell1963excitatory,matzner2000neuromuscular}. 
However, the exact neural circuitry and its functioning for controlling arm movements remain elusive to researchers.

%%%%%%%%%%%%%%%%%%%%%%%%%%%%%%%%%%%%%%%%%%%%%%%%%%%%%%%%%%%%%%%%%%%%%%%%
%%%%%%%%%%%%%%%%%%%%%%%%%%%%%%%%%%%%%%%%%%%%%%%%%%%%%%%%%%%%%%%%%%%%%%%%%

\textit{Musculature.} The ANC is surrounded by densely packed muscles which are broadly divided into three categories -- transverse, longitudinal, and oblique~\citep{kier2007arrangement, kier2016musculature}. Transverse muscles are oriented orthogonally to the axial direction and are responsible for extending the arm. Longitudinal muscles run parallel to the ANC and their contractions bend the arm. Finally, oblique muscles are arranged in a helical fashion, making the arm twist upon contraction.

\noindent
{\bf 2) Behavioral observations}

\noindent
a) {\it Curled rest state:}
Octopus arms tend to curl and form spirals, particularly when at rest \citep{packard1971body, mather1998octopuses}, exposing suckers  outwardly as shown in Fig.~\ref{fig:curl}(a). This curling behavior may be beneficial for several biophysical reasons. Examples range from protecting the arm from predators, environmental awareness owing to large number of exposed suckers, or providing an efficient initial state for arm reaching motions.
%better sensing of the environment since larger proximal suckers face outward, or providing a more efficient initial state for arm reaching motions.
%: i) it protects the arms from damage by predators, ii) it allows the arms to better sense the environment with larger proximal suckers facing outward, iii) it can provide better adhesion to surfaces, particularly when an octopus is resting, as curling can cluster together suckers close to the octopus body, and iv) it may serve as a more efficient initial state for arm reaching motions~\citep{gutfreund1998patterns, sumbre2001control}, where bend propagation initiated from the base of a curled arm can cause the arm to effectively unfurl. On the contrary, more effort would be required if starting from a straightened arm. 
Notably, curling at rest seems to persist even without input from the central nervous system (CNS), as observed in isolated arms (Fig.~\ref{fig:curl}(a)). Mechanically, arm curling at rest may involve inherent tensions of the arm muscle groups~\citep{di2021beyond}. 
%The mechanism for arm curling at rest may involve inherent tensions in the octopus arm muscle groups, such as those shown by \cite{di2021beyond}. %where longitudinal muscles are maintained under 20\% compression, while transverse muscles are maintained under 30\% stretching, at rest.

\medskip
b) {\it Goal-directed arm motions:} One of the most common octopus arm motions involves bend propagation wherein the arm creates a bend at the base and actively propagates it along the length of the arm through traveling waves of muscle actuations~\citep{gutfreund1998patterns, sumbre2001control, hanassy2015stereotypical}. This reaching motion persists in severed arms \citep{sumbre2001control}, suggesting that its motor program could be encoded in the arm PNS as a motion primitive. Fig.~\ref{fig:reaching}(a) showcases video frames from one such reaching motion where the curling at the tip is also visible.

%{\bf 2) Curled shape at rest state:} It has been observed in octopus arm that the rest state is in curled shape

%\noindent
%{\bf 3) Neuromuscular control:}

\subsection{Contributions}
This paper is a continuation of our previous work~\citep{chang2021controlling, wang2022control, wang2022sensory} on control strategies for soft arms modeled via Cosserat rod theory \citep{antman1995nonlinear, gazzola2018forward}. The primary contributions of this work are the following.

\noindent
{\bf 1) Modeling of the neuronal architecture:} A mathematical abstraction of the arm PNS is proposed. Continuum neuronal activity is captured using cable theory~\citep{tuckwell1988introduction} and neuromuscular control is in the form of distributed current inputs to the neurons. The coupling between neural and muscular systems, through muscle innervation, is also modeled. This yields a compact description of the soft arm in a system-theoretic manner.

%anatomy is proposed for octopus neural architecture. 

\noindent
{\bf 2) Analysis of the neuromuscular control system:}

{\it a) Characterizing the rest state:} An analysis of equilibria of the coupled arm dynamics is provided to show that different curled rest shapes can be obtained. 
%the curled rest state of octopus arm and how different curled shapes are obtained. 
A qualitative comparison is provided between numerical results and biological observations. %Besides the curled equilibrium, a set of initial bent configurations for reaching motion is also illustrated.

\noindent
{\it b) Control problem:} Two control laws are proposed for the neuromuscular control system and are shown to accomplish the task of reaching a stationary target. Stability analyses are provided for the same. 

%Based on our previous work on sensor model for octopus arm, a sensory feedback control law is applied to the proposed neuromuscular architecture and stability analysis is given for reaching a stationary target in the workspace.

The rest of the paper is organized as follows. In Sec.~\ref{sec:model}, we introduce the proposed neuromuscular control architecture. In Sec.~\ref{sec:analysis}, we analyze the arm rest state and describe the neuromuscular control laws for the reaching problem. We provide simulation results in Sec.~\ref{sec:simulation} and conclude in Sec.~\ref{sec:conclusion}.
% are demonstrated in Sec.~\ref{sec:simulation}. Sec.~\ref{sec:conclusion} is the conclusion.

\section{Modeling} \label{sec:model}
A neuromuscular octopus arm is comprised of several components, including arm musculature, peripheral nervous system, the coupling between the neural (electrical) and the muscular (mechanical) subsystems, and sensing, as shown in Fig.~\ref{fig:neuromuscular_model}(d). In the rest of this section, we provide models for each of these components.

\subsection{Muscular arm}
A soft octopus arm is modeled as a planar Cosserat rod \citep{antman1995nonlinear, chang2020energy, chang2021controlling}. In this paper, we consider an inextensible and unshearable rod (Kirchhoff rod) for simplicity of exposition. Let $\set{{\mathbf{e}}_1,\mathbf{e}_2}$ denote a fixed orthonormal basis for the two-dimensional laboratory frame. The independent variables are time $t\in\R$ and arc-length $s\in[0,L]$, where $L$ is the length of the rod (see Fig.~\ref{fig:neuromuscular_model}(b)). The subscripts $(\cdot)_t$ and $(\cdot)_s$ denote partial derivatives with respect to $t$ and $s$, respectively. 

The position vector of the centerline is denoted by $\mathbf{r}(s,t) \in \R^2$ and the angle $\theta(s,t) \in[0,2\pi)$ describes the material frame spanned by the orthonormal basis $\set{\mathbf{a}, \mathbf{b}}$, where $\mathbf{a} = \cos \theta \,\mathbf{e}_1 + \sin \theta \, \mathbf{e}_2, ~ \mathbf{b} = -\sin \theta \, \mathbf{e}_1 + \cos \theta \, \mathbf{e}_2$. The kinematics of the rod are given by
\begin{equation}
	\mathbf{r}_s = \begin{pmatrix} \cos \theta \\ \sin \theta \end{pmatrix} = \mathbf{a}, \quad \theta_s = \kappa
	\label{eq:kinematics}
\end{equation}
where $\kappa$ is the curvature of the centerline. The dynamics are described by the set of partial differential equations~\citep{antman1995nonlinear,gazzola2018forward,chang2021controlling}
\begin{equation}
\begin{aligned}
(\varrho A \mathbf{r}_t)_t &= (\mathsf{Q}\mathbf{n})_s - \zeta \mathbf{r}_t + \mathbf{f}^{\text{drag}} \\
(\varrho I \theta_t)_t &= (EI\kappa)_s + n_2  - \zeta \theta_t + u_s 
\end{aligned}
\label{eq:dynamics}
\end{equation}
where $\varrho$ is density, $A$ is the cross sectional area, $\mathsf{Q} = [\mathbf{a} ~~ \mathbf{b} ]$ is a planar rotation matrix, and $\zeta > 0$ is a damping coefficient which models viscoelastic dissipation within the arm. For a linearly elastic arm, $\mathbf{n} = n_1 \mathbf{a} + n_2 \mathbf{b}$ and $EI\kappa$ are internal passive forces and couples, where $E$ is the Young's modulus and $I$ is the second moment of area of the cross section. The effect of drag forces due to the surrounding fluid environment is captured by $\mathbf{f}^{\text{drag}}$, details of which can be found in~\citep{wang2022sensory}. 

In this paper, we only consider two longitudinal muscles (top $(\LMt)$ and bottom $(\LMb)$, see Fig.~\ref{fig:neuromuscular_model}(b)). %These muscles run along the length of the arm and are responsible for bending the arm. 
Since the arm is assumed to be inextensible and unshearable, we simply model muscle actuations as internal couples on the arm.
%without loss of generality we model the muscles to be exerting couples on the rod.
% couple control, denoted by $u$. 
Let $u^{\LMt}$ and $u^{\LMb}$ denote the muscle couples for the top and bottom longitudinal muscles, respectively. Then the total active muscle couple $u$ in \eqref{eq:dynamics} is calculated as $u = u^{\LMb} - u^{\LMt}$. Additional details on muscle modeling are found in \citep{chang2021controlling}. Combined with the passive elastic response, the total internal couple is then $m = EI\kappa + u$. Finally, the dynamics \eqref{eq:dynamics} are accompanied by a fixed-free boundary condition 
\begin{align}
\mathbf{r}(0, t) = 0,~ \theta(0, t) = 0, ~\mathbf{n}(L, t) = 0,~ m(L, t) = 0
\label{eq:arm_boundary_conditions}
\end{align}
 
% Muscle actuations as a function of the couple control $u$ are then given by
%\begin{equation}
%	u^{\LMt} = -u  {\mathds{1}\{u\leq0\}}, \quad
%	u^{\LMb} = u  {\mathds{1}\{u>0\}}
%	\label{eq:distributed_couple}
%\end{equation}
%where $\mathds{1}\{\cdot\}$ is the indicator function.

\subsection{Peripheral nervous system (PNS)}
Muscle contraction control inputs are provided by the underlying electrical activity of neurons. %, which is characterized by the membrane potential or voltage. 
The first mathematical abstraction for an excitable neural media was given by Hodgkin and Huxley~\citep{hodgkin1952quantitative} to describe how the membrane potential or voltage evolves and propagates in a squid giant axon. Along with the membrane potential, the Hodgkin-Huxley model contains three other variables to capture the effects of sodium (Na$^+$), potassium (K$^+$), and leakage currents in the action potential dynamics. In the following decades, a wide range of simplified mathematical models were proposed, a summary of which can be found in~\citep{izhikevich2007dynamical}.

The musculature model in this paper considers two longitudinal muscles independently controlled by the PNS, whose central component is the axial nerve cord (ANC). We model the ANC by two long cylindrical nerves, referred to here as `cables' (see Fig.~\ref{fig:neuromuscular_model}(c)). We model the neural activity of these nerves after the cable theory~\citep{rall1962theory, tuckwell1988introduction} which is a simplified model for one-dimensional excitable media.
%We adopt a simplified model for one-dimensional excitable media, known as the cable theory~\citep{rall1962theory, tuckwell1988introduction}, to describe the neural activity of these cables. 
Each of the two cables is characterized by its distributed membrane voltage $\Vmuscle (s,t)$ and a recovery or adaptation variable $\VmA (s,t)$, where $\muscle\in \{\LMt, \LMb\}$. The dynamics of such a cable are expressed as the following partial differential equations
\begin{subequations}
	\begin{align}
	\left(\tau { \Vmuscle} \right)_t &= \lambda^2 \Vmuscle_{ss} - \Vmuscle - \VmA + \Imuscle \label{eq:cable_eq_V} \\
	\left( \tauAdapt { \VmA}\right)_t &= -\VmA + bg(\Vmuscle)
	\end{align}
	\label{eq:cable_eq_full}
\end{subequations}
where $\tau$ is a time constant, $\lambda$ is a length constant, $\tauAdapt$ is the adaptation rate, and the function $g(x)=\max(x,0)$ describes the output property of neurons (ReLU). 
The recovery variable $\VmA$ captures the effect of neuron `fatigue' or self-inhibition and $b$ is the adaptation strength~\citep{matsuoka1984dynamic}. Intuitively, the adaptation decreases the effective length constant leading to a faster decaying (in $s$) neural activity (see discussions in Sec.~\ref{sec:equilibrium_configuration}). Finally, the term $\Imuscle(s,t)$ denotes the total stimulus current which are produced by the local neural circuitry contained in the ganglia. Since the details of such circuitries are not fully known, we model their net effect by the current inputs to the cables. Therefore the variables $I^\muscle$ are considered to be the control inputs for the arm neuromuscular control system (Fig.~\ref{fig:neuromuscular_model}(d)). Context dependent boundary conditions for the voltage equation~\eqref{eq:cable_eq_V} are discussed in Sec.~\ref{sec:analysis}.

%The one-dimensional continuum electrical dynamics of the nerve are represented by cable theory~\citep{rall1962theory, tuckwell1988introduction, brzychczy2013mathematical}.

% Since the two longitudinal muscles are independently controlled, here we consider two long cylindrical nerves known as `cables' \citep{tuckwell1988introduction, brzychczy2013mathematical} which run along the length of the arm (see Fig.~\ref{fig:neuromuscular_model}(c)). Let us denote the local membrane potential of these cables by $V^\muscle(s,t)$ where $\muscle \in \{\LMt, \LMb\}$. As a simplification of the original Hodgkin-Huxley model~\citep{hodgkin1952measurement}, the dynamics of a continuum excitable media are expressed by the following partial differential equations % ~\citep{blah}

%where $\VmA$ is the recovery or adaptation variable, $\tau$ is a time constant and $\tauAdapt$ is the adaptation rate, $b$ is the adaptation strength, $\lambda$ is a length constant, and the function $g(x)=\max(x,0)$ describes the output property of neurons (ReLU). 

%\udit{Not sure if the last line of the previous paragraph is needed?}

%The voltage equation~\eqref{eq:cable_eq_V} is typically accompanied with a closed-open type of boundary conditions
%\begin{align*}
%\Vmuscle (0,t) = \Vmuscle_0 and \Vmuscle_s (L, t) = 0, 
%\end{align*}   
%although different boundary conditions will arise 

\begin{remark}
Notice that on the right hand side of~\eqref{eq:cable_eq_V}, the linear term $(-\Vmuscle)$ is chosen here for simplicity~\citep{matsuoka1984dynamic}. However, other forms have been studied and can be incorporated into our model. For example, the Fitzhugh-Nagumo model~\citep{fitzhugh1961impulses, nagumo1962active} uses a cubic term $\Vmuscle - \tfrac{1}{3}(\Vmuscle)^3$. Moreover, mutual inhibition has also been studied for models of central pattern generators in octopus arms~\citep{tian2015simulation}, a topic we postpone for future studies.  
 %and can be incorporated in our model. A notable example is the Fitzhugh-Nagumo model~\citep{fitzhugh1961impulses, nagumo1962active} where a cubic term $\Vmuscle - \tfrac{1}{3}(\Vmuscle)^3$ is instead used.  
\end{remark}

\begin{remark}
In cable theory, the length constant $\lambda$ depends on the radius of the cable \citep{koch1984cable}. However, measurements in octopus suggest that the cross sectional area of the nerve cord does not decrease significantly, even though the arm tapers toward the tip~\citep{kier2007arrangement}. For this reason, we take $\lambda$ to be a constant. %in this study.  
\end{remark}

\subsection{Coupling -- from neural activity to muscle contractions}

The neural activity or the cable voltages are responsible for instructing the muscles to contract through motorneuron innervations. The unique neuromuscular system of the octopus arm shows the lack of short-term synaptic plasticity and postsynaptic inhibition in the neuromuscular junction, which suggests an approximately linear transformation of motor neuronal activity into muscular activation~\citep{matzner2000neuromuscular,nesher2020octopus}. We model the coupling between the voltage level and the resulting muscle couple using an activation function $\sigma(\cdot) : \R \mapsto [0, 1]$ as follows
\begin{equation}
	u^\muscle (s,t) = c(s) \sigma(V^\muscle (s,t))
	\label{eq:neuromuscular_mapping}
\end{equation}
where $c(s)>0$ is the maximum couple a muscle can exert along the arm. The activation function $\sigma(\cdot)$ behaves like a saturation function, i.e. it remains close to zero for low voltages, then increases approximately linearly, and finally saturates to one for high voltages, reflecting the physical limits of muscles. The explicit formula for $\sigma(\cdot)$ is provided in Sec.~\ref{sec:simulation}. Further details of the excitation-contraction dynamics for skeletal muscles can be found in~\citep{hatze1977myocybernetic, audu1985influence}.

%\begin{remark}
%Details of the excitation-contraction dynamics for skeletal muscles can be found in 
%\end{remark}

%\udit{Is $c_{\max}$ a function of $s$? If so, indicate in this section and explain in the simulation setup section.}

\subsection{Sensing}
Biological experiments suggest that octopuses use various sensory information, ranging from visual, chemical to proprioceptory, to carry out various manipulation tasks, including reaching to a target~\citep{gutnick2020use, mather2021octopus}. %citations needed
Based on these evidences, we introduced a simplified sensory model in our recent work~\citep{wang2022sensory}. The sensory model assumes that the arm has access to the bearing angle $\alpha$ and the location ($s = \bar{s}$) of the arm closest to the target (see Fig.~\ref{fig:neuromuscular_model}(b)). Further details of the sensing model can be found in~\citep[Sec.~III-A.]{wang2022sensory}.

\section{Analysis of the neuromuscular control system} \label{sec:analysis}

\subsection{Characterizing the rest shape} \label{sec:equilibrium}

\begin{figure*}[t]
	\centering
	\includegraphics[width=\textwidth, trim = {0pt 0pt 0pt 0pt}]{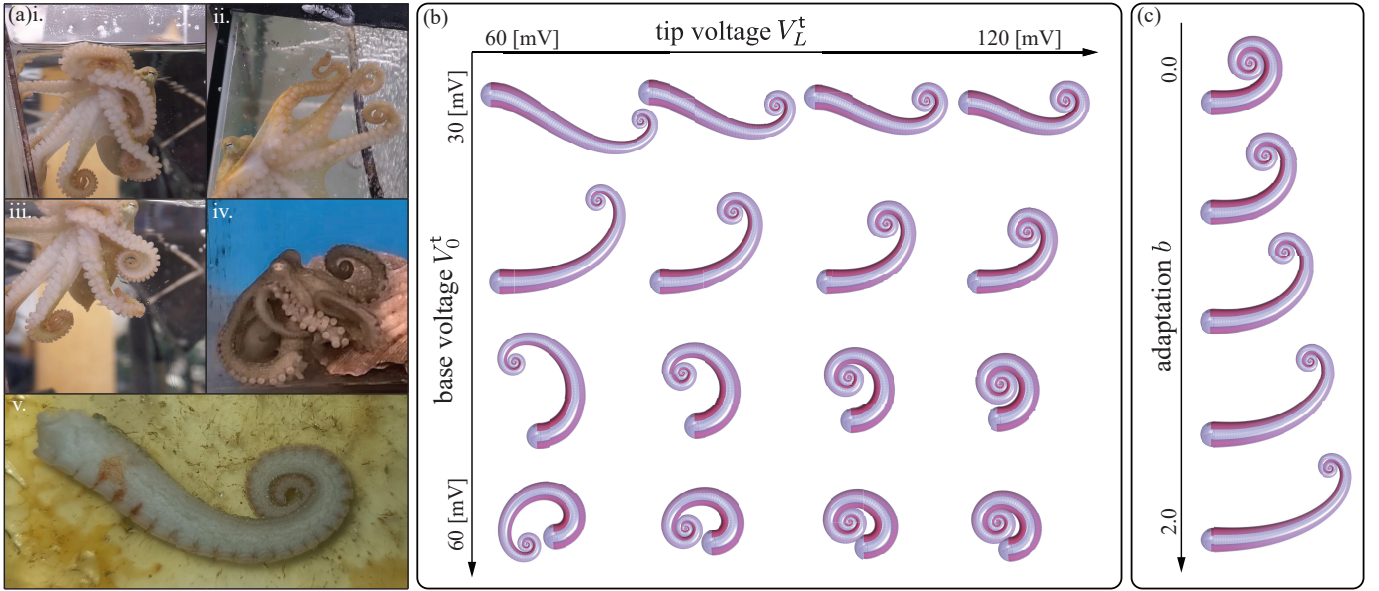}
	\caption{Curled shapes in both real octopus arms and simulated arms. (a) Snapshots of different octopuses showing curled arm at rest. Subfigures (i.-iv.) are of live octopuses while the subfigure v. is an severed arm that remains curled even after isolation. (b) The equilibrium of the arm under varying base and tip voltages $\Vtop_0$ and $\Vtop_L$ for the top longitudinal muscle. The intensities of the distributed muscle actuation are illustrated in red. (c) The equilibrium of the arm under varying adaptation parameter $b$.}
	\label{fig:curl}
	\vspace{0pt}
\end{figure*}

In this section, we are interested in explaining the curled rest shapes as observed both in attached and isolated arms (Fig.~\ref{fig:curl}(a)), which is suggestive of tonic muscle activity. To incorporate such biological observations within our framework, we first study the equilibrium of the cable equations \eqref{eq:cable_eq_full} with a fixed-fixed boundary condition under zero control input~{$(I^\muscle = 0)$}. %and a fixed-fixed boundary condition. 

%It is commonly observed in octopus arm that its rest state forms a curled shape (see Fig.~\ref{fig:curl}(a), we are interested in explaining the curled rest state of the arm. Biologist studies point out that the muscles may remain tense even at rest state~\cite{di2021beyond}. We reproduce such property by incorporating a fixed-fixed boundary condition with zero control input.

The equilibrium voltage of the cables must satisfy %the equations of statics obtained from the dynamics of the neurons~\eqref{eq:cable_eq_full} as:
\begin{equation}
		\lambda^2 \Vmuscle_{ss} = \Vmuscle + bg(\Vmuscle)
	\label{eq:equilibrium_no_control}
\end{equation}
with fixed-fixed boundary conditions
\begin{equation}
		\Vmuscle\Big|_{s=0} = \Vmuscle_0,\ \Vmuscle\Big|_{s=L} = \Vmuscle_L
	\label{eq:voltage_boundary_conditions}
\end{equation}
where $\Vmuscle_0$ and $\Vmuscle_L$ are the fixed voltages for muscle $\muscle$ at the base and at the tip, respectively. Depending on the signs of the boundary voltages, we can explicitly express the equilibrium voltage as presented next.
%solutions to ~\eqref{eq:equilibrium_analytical_solution}. 
%We present this as the following proposition. % whose proof appears in Appendix~\ref{appdx:equilibrium_V_proof}. 

%Solutions to \eqref{eq:equilibrium_no_control} are expressed as a sum of two exponential functions as follows
%\begin{align}
%\Vmuscle (s) = c_1^\muscle e^{\tfrac{s}{\hat{\lambda}}} + c_2^\muscle e^{-\tfrac{s}{\hat{\lambda}}}
%\label{eq:V_equilibrium}
%\end{align}
%where $c_1^\muscle, c_2^\muscle, \hat{\lambda}$ are constants which depend on $\Vmuscle_0, \Vmuscle_L, \lambda$ and $b$. The derivation of \eqref{eq:V_equilibrium} is straightforward and is omitted here due to lack of space.

%, and $\tilde{\lambda}$ takes value from $\{\tfrac{\lambda}{\sqrt{1+b}}, \lambda \}$ depending on $\Vmuscle_0$ and $\Vmuscle_L$.

\begin{lemma} 
%Let us denote $\tilde{\lambda}= \frac{\lambda}{\sqrt{1+b}}$. Then,
Consider the voltage equilibrium equation \eqref{eq:equilibrium_no_control} with boundary conditions \eqref{eq:voltage_boundary_conditions}.  %Without loss of generality, we only present solutions for the following cases.

a) 	If $\Vmuscle_0 \Vmuscle_L \geq 0$, then %the solution to \eqref{eq:equilibrium_no_control} is given by
\begin{align}
\Vmuscle (s) = c_1^\muscle e^{\tfrac{s}{\hat{\lambda}}} + c_2^\muscle e^{-\tfrac{s}{\hat{\lambda}}} := f(s; c_1^\muscle, c_2^\muscle, \hat{\lambda})
\end{align}
where $c_1^\muscle, c_2^\muscle, \hat{\lambda}$ are constants which depend on $\Vmuscle_0, \Vmuscle_L, \lambda$ and $b$. %, and $\tilde{\lambda}$ takes value from $\{\tfrac{\lambda}{\sqrt{1+b}}, \lambda \}$ depending on $\Vmuscle_0$ and $\Vmuscle_L$.

b) If $\Vmuscle_0 \Vmuscle_L < 0$, then
\begin{align}
\Vmuscle (s) = \begin{cases}
f(s; k_1^\muscle, k_2^\muscle, \hat{\lambda}_1), \quad 0 \leq s \leq s_1^\muscle \\
f(s; p_1^\muscle, p_2^\muscle, \hat{\lambda}_2), \quad s_1^\muscle \leq s \leq L
\end{cases}
%\Vmuscle (s) = \begin{cases}
%k_1^\muscle e^{\tfrac{s}{\hat{\lambda}_1}} + k_2^\muscle e^{-\tfrac{s}{\hat{\lambda}_1}}, \quad 0 \leq s \leq s_1^\muscle \\
%p_1^\muscle e^{\tfrac{s}{\hat{\lambda}_2}} + p_2^\muscle e^{-\tfrac{s}{\hat{\lambda}_2}}, \quad s_1^\muscle \leq s \leq L
%\end{cases}
\end{align}
where $k_1^\muscle, k_2^\muscle, p_1^\muscle, p_2^\muscle, \hat{\lambda}_1, \hat{\lambda}_2$ are constants which depend on $\Vmuscle_0, \Vmuscle_L, \lambda$ and $b$. Furthermore, the zero-crossing point $s_1^\muscle$ is found by solving the following nonlinear equation 
\begin{align*}
f_s (s_1; k_1^\muscle, k_2^\muscle, \hat{\lambda}_1) = f_s (s_1; p_1^\muscle, p_2^\muscle, \hat{\lambda}_2))
%\hat{\lambda}_2 \left( k_1^\muscle e^{\tfrac{s_1^\muscle}{\hat{\lambda}_1}} - k_2^\muscle e^{-\tfrac{s_1^\muscle}{\hat{\lambda}_1}} \right) = \hat{\lambda}_1\left(p_1^\muscle e^{\tfrac{s_1^\muscle}{\hat{\lambda}_2}} - p_2^\muscle e^{-\tfrac{s_1^\muscle}{\hat{\lambda}_2}}\right)
\end{align*}
\label{lemma:equilibrium_no_control}
\end{lemma}
The proof of Lemma~\ref{lemma:equilibrium_no_control} is straightforward and utilizes the rectified linear property of the output function $g(\cdot)$. The proof is omitted here due to lack of space. 
%With the equilibrium voltage of motor neurons~\eqref{eq:equilibrium_analytical_solution}, 

We next describe how these equilibrium voltages affect the shape of the arm. A straightforward calculation of the statics of~\eqref{eq:dynamics} with the boundary conditions~\eqref{eq:arm_boundary_conditions} yields the following expression for equilibrium curvature
%. the equilibrium curvature of the arm must satisfy
\begin{equation}
	\kappa(s) = \frac{c(s)}{EI(s)}\left(\sigma(\Vtop) - \sigma(\Vbot) \right) %+ \bar{\kappa}
	\label{eq:equilibrium_kappa}
\end{equation}
%where $\bar{\kappa}$ is a constant of integration that depends upon the boundary voltage values. 
%Equation~\eqref{eq:equilibrium_kappa} is obtained by considering the statics of  ~\eqref{eq:dynamics} with the boundary conditions~\eqref{eq:arm_boundary_conditions}. 
Once the curvature $\kappa$ is found, the configuration of the arm is obtained by the virtue of the kinematics~\eqref{eq:kinematics}.
%We vary different parameters to investigate how the equilibrium configuration of the arm changes. 
The equilibrium shapes are obtained numerically as discussed in Sec.~\ref{sec:equilibrium_configuration}.
% illustrated in Fig.~\ref{fig:curl} and 

%The equilibrium of the voltage must satisfy the equations of statics obtained from the dynamics of the neurons~\eqref{eq:cable_eq_full} as:
%\begin{equation}
%	\begin{aligned}
%		\lambda^2 \Vmuscle_{ss} &= \Vmuscle + bg(\Vmuscle)
%	\end{aligned}
%	\label{eq:equilibrium_no_control}
%\end{equation}
%with the given fixed boundary conditions
%\begin{equation}
%	\begin{aligned}
%		\Vmuscle\Big|_{s=0} = \Vmuscle_0,\ \Vmuscle\Big|_{s=L} = \Vmuscle_1
%	\end{aligned}
%	\label{eq:voltage_boundary_conditions}
%\end{equation}
%
%Under the equilibrium voltage of motor neurons, an equilibrium of the arm can be found which satisfies the statics of the dynamics~\eqref{eq:dynamics} and the boundary conditions~\eqref{eq:arm_boundary_conditions} by solving the following equation for the curvature $\kappa$
%\begin{equation}
%	\kappa(s) = \frac{-1}{EI(s)}\left(\sigma(\Vtop) - \sigma(\Vbot) \right)
%	\label{eq:equilibrium_kappa}
%\end{equation}
%
%We vary different parameters to investigate how the equilibrium configuration of the arm changes. The equilibrium solutions are obtained numerically, then illustrated and discussed in Fig.~\ref{fig:curl} and Sec.~\ref{sec:equilibrium_configuration}.

\subsection{The reaching problem} \label{sec:ctrl_prop}

In this section, we study the dynamic control problem of designing the input currents $I^\muscle(s,t)$ to drive the arm towards a given static target located at $\mathbf{r}^{\text{target}}$ (Fig.~\ref{fig:neuromuscular_model}(b)). The reaching task is considered completed when the arm stabilizes and either `touches' (target within reach) or `points toward' (target out of reach) the target. A formal definition of these two scenarios are found in \citep{wang2022sensory}. In contrast to the fixed-fixed boundary condition for the cable equation \eqref{eq:cable_eq_V} under zero control input for rest state configuration, here we use a free-free boundary condition 
\begin{equation}
	\Vmuscle_s \Big|_{s=0} = \Vmuscle_s \Big|_{s=L} = 0
	\label{eq:voltage_boundary_conditions_with control}
\end{equation}
which may be explained by the octopus CNS `switching' from the rest state to the motile state. We next discuss two current control laws.

\textbf{1) Reference tracking control:}  
Suppose a reference muscle couple control $\bar{u}^\muscle = c(s)\bar{v}^\muscle(s,t), ~ \bar{v}^\muscle \in [0,1]$, is given which solves the reaching problem. %and (locally) stabilizes the arm dynamics~\eqref{eq:dynamics}. 
Such a stabilizing control may be obtained by various control methods, e.g. the energy shaping control~\citep{chang2020energy, chang2021controlling} or feedback control~\citep{wang2022sensory}. 
The goal for the neural system is to find current control inputs $I^\muscle(s,t)$ which generate -- via the cable dynamics and the neuromuscular coupling -- muscle actuations that track the reference couple and effectively stabilize the arm. Here we propose such a control current as 
%that takes the feedback on voltage given by
\begin{equation}
\small
		\Imuscle = bg(\Vmuscle) + (1-\beta)\Vmuscle + \beta\sigma^{-1}(\bar{v}^{\muscle}) - \lambda^2\left(\sigma^{-1}(\bar{v}^{\muscle})\right)_{ss}
	\label{eq:reference_tracking_control}
\end{equation}
%\begin{equation}
%	\begin{aligned}
%%		\Imuscle &= \tau\nabla_{\bar{u}^\muscle}\left[\sigma^{-1}(\bar{u}^{\muscle})\right]u^{\muscle}_t + \gamma\tau\sigma^{-1}(u^{\muscle})  \\ & ~~~ + (1-\gamma\tau)\Vmuscle - \lambda^2\Vmuscle_{ss} + \VmA
%	\Imuscle =& \gamma\tau\sigma^{-1}(\bar{u}^{\muscle}) + (1-\gamma\tau)\Vmuscle - \lambda^2\Vmuscle_{ss} + bg(\Vmuscle)
%	\end{aligned}
%	\label{eq:refernece_tracking_control}
%\end{equation}
where $\beta>0$ is a constant. Notice that the activation function $\sigma(\cdot)$ is monotonous so the inverse $\sigma^{-1}(\cdot)$ exists. We refer to the control law \eqref{eq:reference_tracking_control} as reference tracking control whose stability properties are discussed next.

%%%%%%%%%%%%%%%%%%%%%%%%%%%%%%%%%%%%%%%%%%%%%%%%%%%%%%%%%%%%%%%%%%%%%%%%%
%%%%%%%%%%%%%%%%%%%%%%%%%%%%%%%%%%%%%%%%%%%%%%%%%%%%%%%%%%%%%%%%%%%%%%%%%

\begin{proposition}
	Suppose the muscular arm~\eqref{eq:dynamics} is (locally) asymptotically stable under the reference couple control $\bar{u}^\muscle$. Then, there exists a $\beta >0$ large enough so that the neuromuscular arm~(\ref{eq:dynamics},\,\ref{eq:cable_eq_full},\,\ref{eq:neuromuscular_mapping}) is also (locally) asymptotically stable under the current control~\eqref{eq:reference_tracking_control}.
%	Consider the equilibrium that is (locally) asymptotically stable to the system~\eqref{eq:dynamics} under the benchmark control $u^\star$. Then, the same equilibrium is also (locally) asymptotically stable to the neuromuscular arm~\eqref{eq:dynamics}\eqref{eq:cable_eq_full}\eqref{eq:neuromuscular_mapping} with the feedback control law \eqref{eq:backstepping_control}.
	\label{prop:backstepping}
\end{proposition}

%\udit{Write a better statement of the proposition.}

The proof of the proposition relies on separating the time scales of the electrical and mechanical subsystems \citep[Chap. 11.5]{khalil2002nonlinear}. A sketch of the proof is given in Appendix~\ref{appdx:backstepping_proof}. %as follows:
%
%\udit{End this subsection by discussing the pros and cons of this control law. Pro: you can use any control law $u$ as the reference, not just the $u^\star$ and the neural system tracks that reference. Con: It requires feedback on the voltage which is unrealistic. For this reason, we turn to the next control law.}
The benefit of the control law~\eqref{eq:reference_tracking_control} is that any muscle control law can be used as a reference with the neural system being able to track it. However, it is biologically unrealistic to have feedback on the cable voltage. For this reason, we next turn to a more realistic control law. 

%\subsection{Sensory feedback control} \label{sec:sensory_feedback_control}

\textbf{2) Sensory feedback neural control:}
Given the sensory information $(\alpha, \bar{s})$, we propose a feedback control law at the neural level as 
\begin{align}
	\begin{split}
	&I(s,t) = -\mu \sin(\alpha(s,t))\mathds{1}\{s\leq\bar{s} (t)\}, ~~ \mu > 0 \\
	&\Itop = -I \mathds{1} \{I\leq0\}, ~~ \Ibot = I{\mathds{1}\{I>0\}}
	\end{split}
	\label{eq:sensory_feedback_control_law}
\end{align}
%where $\mu$ is a constant, $\alpha$ and $\bar{s}$ are the sensory information (see Appendix~\ref{appdx:sensor_model} for details). 
% The current inputs for top and bottom longitudinal muscles as a function of the above total current $I$ are given by
%\begin{equation}
%	\Itop = -I{\mathds{1}\{I\leq0\}}, \quad \Ibot = I{\mathds{1}\{I>0\}}
%	\label{eq:sensory_feedback_control_law_distributed}
%\end{equation}
%Even though essentially the same form of the benchmark control is used at the neural level, the closed-loop system remains stable while accomplishing the control task. This is described by the following proposition.
The control law makes the arm active up to the $\bar{s}$ point and passive afterwards. The motivations behind such a feedback control law was the main contribution of our previous work \citep{wang2022sensory}. In particular, a central result in~\citep[Proposition 4.1]{wang2022sensory} states that a couple control law with the same form of the right hand side of \eqref{eq:sensory_feedback_control_law} achieves the goal of reaching. In this paper, we refer to such a couple control law as the benchmark and denote it by $u^\star$.

Next we discuss the stability properties of the feedback neural control~\eqref{eq:sensory_feedback_control_law}.

\begin{proposition}
Consider the dynamics of the neuromuscular arm~(\ref{eq:dynamics},\,\ref{eq:cable_eq_full},\,\ref{eq:neuromuscular_mapping}) with the proposed feedback control law \eqref{eq:sensory_feedback_control_law}. Then, there exist a $\mu>0$ large enough so that the feedback control \eqref{eq:sensory_feedback_control_law} accomplishes the reaching task.
	%Consider the dynamics of the neuromuscular arm~\eqref{eq:dynamics}\eqref{eq:cable_eq_full}\eqref{eq:neuromuscular_mapping} with the proposed feedback control law \eqref{eq:sensory_feedback_control_law}. Then the equilibrium derived from~\eqref{eq:equilibrium_kappa} and \eqref{eq:equilibrium_feedback_control} is (locally) asymptotically stable.
	\label{prop:equilibrium}
\end{proposition}

%\udit{Clean up the statement and put all the proof in the appendix.}
The proof of this proposition is similar to that of Proposition~\ref{prop:backstepping} where the sensory feedback control is considered a constant on the fast time-scale of the neural system. The full proof is omitted here on account of space.

\section{Simulation results} \label{sec:simulation}

%%%%%%%%%%%%%%%%%%%%%%%%%%%%%%%%%%%%%%%%%%%%%%%%%%%%%%%%%%%%%%%%%%%%%
\begin{table}[!t]
	\centering
	\caption{Parameters for models and numeric simulation}
	\begin{tabular}{ccc}
		\hline
		\hline\noalign{\smallskip}
		Parameter & Description & Numerical value \\
		\hline\noalign{\smallskip}
		\multicolumn{3}{c}{{\bf Rod model}}\\
		$L$ & length of the undeformed rod [cm] & $20$ \\
		$\rodbase$ & rod base radius [cm] & $1$ \\
		$\rodtip$ & rod tip radius [cm] & $0.1$ \\
		$\varrho$ & density [kg/${\text{m}}^3$] & $1042$ \\
		$\zeta$ & damping coefficient [kg/s]  & $0.01$\\
		$E$ & Young's modulus [kPa] & $10$ \\
		%$\Delta t$ & time step size [s] & $10^{-5}$\\
		\hline\noalign{\smallskip}
		\multicolumn{3}{c}{{\bf Neuron model}}\\
		$\tau$ & time constant [s] & $0.04$ \\
		$\tauAdapt$ & adaptation rate [s] & $0.4$ \\
		$\lambda$ & length constant [cm] & $10$ \\
		\hline 
	\end{tabular}
	\label{tab:num_para}
	\vspace*{0pt}
\end{table}
%%%%%%%%%%%%%%%%%%%%%%%%%%%%%%%%%%%%%%%%%%%%%%%%%%%%%%%%%%%%%%%%%%%%%

In this section, we show numerical simulations of an octopus arm equipped with the proposed neuromuscular control architecture. The arm dynamics \eqref{eq:dynamics}-\eqref{eq:arm_boundary_conditions} are solved using the open-source software \textit{Elastica}~\citep{gazzola2018forward, zhang2019modeling}. In all our simulations, a tapered arm geometry is considered based on measurements in real octopuses~\citep{chang2020energy}.
%In all simulations, the variable $\radius(s) = \rodtip s/L + \rodbase(1 - s/L)$ gives the radius profile of a tapered rod, based on measurements of real octopuses, %The mass-inertia density matrix is given by $\mathcal{M}=\diag(\rho A, \rho A, \rho I)$. $\rho$ is the material density of the rod, 
%$A (s) =\pi(\radius(s))^2$ and $I(s)=\frac{A(s)^2}{4\pi}$ are the cross sectional area and second moment of area, respectively.
%The length constant $\lambda(s)=\lambda_0\sqrt{\frac{\radius(s)}{\rodbase}}$ is a function of the arc-length $s$ and $\lambda_0\in\R^+$. 
The neural dynamics \eqref{eq:cable_eq_full} are numerically solved consistently with the temporal and spatial discretization of \textit{Elastica} for suitable assimilation.
%second order Runge-Kutta method with the same temporal and spatial discretization as in Elastica for suitable assimilation. 
The values of the time constants $\tau, \tilde{\tau}$ are taken in the range identified in \citep{ekeberg1993combined}. 
The activation function in the neuromuscular coupling equation~\eqref{eq:neuromuscular_mapping} is chosen as %such that the value of muscle activation $u$ approches the upper bound $1$ when the voltage $\Vmuscle$ exceeds $80$ [mV]:
\begin{equation}
\sigma(V) = \tfrac{1}{2} \left( 1 + \tanh\left(-\tfrac{1}{40}\arctanh(-0.98(V-40)) \right) \right)
\label{eq:neuromuscular_mapping_explicit}
\end{equation}
whereas the details of the maximum muscle couple coefficient $c(s)$ in~\eqref{eq:neuromuscular_mapping} are given in~\citep{chang2021controlling}. %which describe the properties of the muscles. 
The rest of the parameter values used in simulations are reported in Table~\ref{tab:num_para} and further details and explanations are found in \citep{chang2020energy, chang2021controlling}.
%The constant $\gamma=2500$ in the backstepping control law~\eqref{eq:backstepping_control} and the constant $\mu=500$ in the sensory feedback control law~\eqref{eq:sensory_feedback_control_law} The rest of the parameter values used in simulations are reported in Table~\ref{tab:num_para}.

%%%%%%%%%%%%%%%%%%%%%%%%%%%%%%%%%%%%%%%%%%%%%%%%%%%%%%%%%%%%%%%%%%%%%%%%%
\begin{figure*}[!t]
	\centering
	\includegraphics[width=\textwidth, trim = {0pt 1250pt 0pt 0pt}, clip=true]{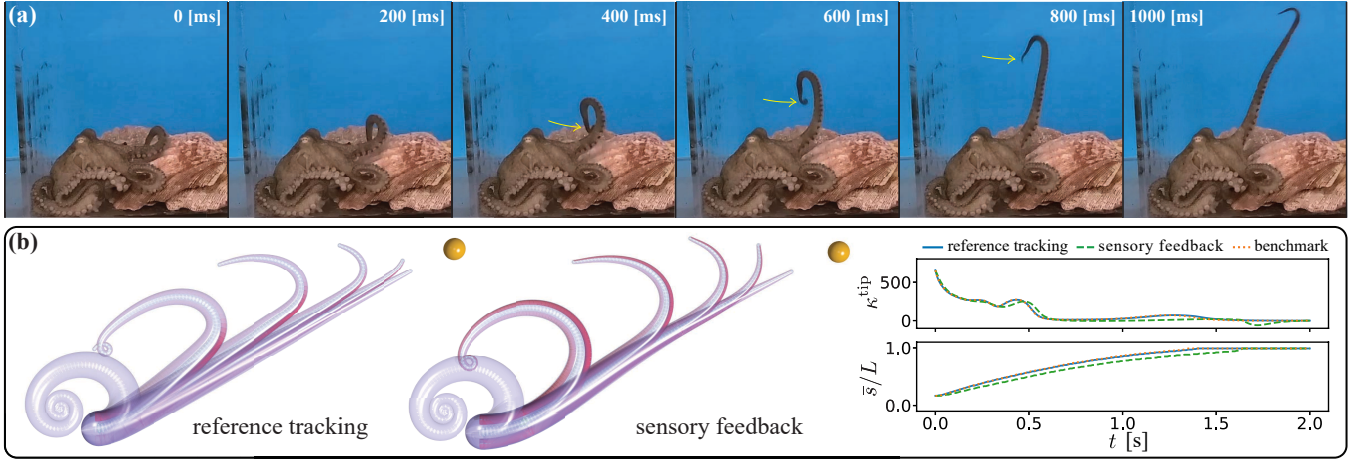}
	\caption{(a) A video sequence of an \textit{Octopus bimaculoides} performing bend propagation. Yellow arrows indicate the initial curling at the tip which gradually vanishes during the bend propagation movement. (b) The rod is initialized with a bend at the base and the curl at the tip. The first and second columns are simulation results of rod reaching towards a static target under reference tracking control and sensory feedback control, respectively. The third column is the $\Ktip$ and $\bar{s}$ trajectories comparison among reference tracking control (blue solid lines), sensory feedback control (green dash lines), and the benchmark control (orange dotted lines). %(c) Same simulation results and trajectories comparison as (b) except for a different rod initialization. The initial rod configuration is curled at the tip but straight at the base. %\udit{Fix the figure.}
	}
	\label{fig:reaching}
	\vspace{0pt}
\end{figure*}
%%%%%%%%%%%%%%%%%%%%%%%%%%%%%%%%%%%%%%%%%%%%%%%%%%%%%%%%%%%%%%%%%%%%%%%%%

\subsection{Rest shape} \label{sec:equilibrium_configuration}

%To obtain the rest state of the arm under zero control input, we initialize the rod to be straight and run the simulation with zero current until the rod is stabilized. 
We obtain the rest state of the arm under zero control input by numerically solving the equilibrium equations~\eqref{eq:equilibrium_no_control}, \eqref{eq:voltage_boundary_conditions}, \eqref{eq:equilibrium_kappa}, and \eqref{eq:kinematics}, as described in Sec.~\ref{sec:equilibrium}. Here we display a collection of rest shapes by varying the boundary voltage values ($\Vmuscle_0, \Vmuscle_L$) and the adaptation parameter $b$.% to get different curled rest states. Second, we vary the adaptation parameter to see how adaptation affects the equilibrium.

We first fix the boundary voltages $\Vbot_0=40$ [mV] and $\Vbot_L=0$ [mV] for the bottom longitudinal muscle. Then, two sets of experiments are carried out.

\noindent
\textit{1) Varying the top boundary voltages:} We fix the adaptation parameter $b=1.0$ and vary the boundary voltages $\Vtop_0$ and $\Vtop_L$ for the top longitudinal muscle by taking their values from the sets $\{30,40,50,60\}$ [mV] and $\{60,80,100,120\}$ [mV], respectively. This results in a 4-by-4 chart as illustrated in Fig.~\ref{fig:curl}(b).
% $\Vtop_0$ takes value among $[30,40,50,60]$ [mV] and $\Vtop_L$ takes value among $[60,80,100,120]$ [mV], which results in a 4-by-4 chart as illustrated in the first panel of Fig.~\ref{fig:curl}(b). 
We observe that as $\Vtop_0$ increases (columns), the arm tends to bend more at the base. On the other hand, as $\Vtop_L$ increases (rows), the tip curls more. Some of the simulated rest shapes present a good qualitative match with the recordings from real octopuses. For instance, the shape in row-2/column-4 matches Fig.~\ref{fig:curl}(a)i.-iii., where the arm is straight near the base and curls up at the tip. Similarly, the row-4/column-2 result shows a tip curling with a bent base, matching the recording Fig.~\ref{fig:curl}(a)iv. Finally, row-1/column-4 presents a shape similar to the isolated curled arm of Fig.~\ref{fig:curl}(a)v. 

\noindent
\textit{2) Varying the adaptation parameter:} We fix the base and tip voltage of the top longitudinal muscle to $\Vtop_0=40$ [mV] and $\Vtop_L=80$ [mV]. Then, we vary the adaptation parameter $b$ from the set $\{0,0.5,1.0,1.5,2.0\}$ and show the resulting rest shapes in Fig.~\ref{fig:curl}(c).
%The adaptation parameter $b$ takes value among $[0,0.5,1.0,1.5,2.0]$. 
We see that as the adaptation increases, the arm tends to unwind, or in other words, the arm curls less. %The effect of adaptation can be directly explained by the fatigue effect of neurons from self-inhibition~\cite{matsuoka1984dynamic}. 
Mathematically speaking, adaptation parameter $b$ takes effect in changing the equivalent length constant $\lambdaA=\frac{\lambda}{\sqrt{b+1}}$. Higher adaptation results in smaller equivalent length constant, hence giving equilibrium voltage solution with faster decaying. 
%~\eqref{eq:equilibrium_analytical_solution} with faster decaying.

%%%%%%%%%%%%%%%%%%%%%%%%%%%%%%%%%%%%%%%%%%%%%%%%%%%%%%%%%%%%%%%%%%%%%%%%%%
%\begin{figure}[t]
%	\centering
%	\includegraphics[width=\columnwidth, trim = {0pt 0pt 0pt 0pt}]{Figures/curvature_compare.png}
%	\caption{Comparison of curvatures between the backstepping control result (blue solid lines) and the sensory feedback control results in~\cite{wang2022sensory} (orange dash lines).}
%	\label{fig:backstepping}
%	\vspace{0pt}
%\end{figure}
%%%%%%%%%%%%%%%%%%%%%%%%%%%%%%%%%%%%%%%%%%%%%%%%%%%%%%%%%%%%%%%%%%%%%%%%%%

\subsection{Reaching maneuver} \label{sec:arm_movements}
Here we demonstrate goal-directed arm movements as a solution to the reaching problem described in Sec.~\ref{sec:ctrl_prop}. %We also applied the sensory feedback control law \eqref{eq:benchmark} at muscle actuation level in~\cite{wang2022sensory} as the benchmark. 
%Two sets of experiments are carried out. 
For this experiment, the arm is initialized at the equilibrium configuration under voltage boundary values $\Vtop_0=65$ [mV], $\Vtop_L=65$ [mV], $\Vbot_0=40$ [mV], $\Vbot_L=0$ [mV], and the adaptation parameter $b=1.0$. This initial configuration corresponds to row-4/column-2 shape in Fig.~\ref{fig:curl}(b). A static target at coordinates $[0.2,0.1]$ [m], outside the arm's reach, is presented. The control gains $\beta = 10$ and $\mu = 500$ are chosen for reference tracking control law~\eqref{eq:reference_tracking_control} and sensory feedback control law~\eqref{eq:sensory_feedback_control_law}, respectively. For reference tracking, the benchmark control $u^\star$ is chosen as the reference. 
In Fig.~\ref{fig:reaching}(b), we illustrate arm simulations under both reference tracking control~\eqref{eq:reference_tracking_control} and sensory feedback neural control~\eqref{eq:sensory_feedback_control_law}. As we can see from the snapshots of the dynamic simulation in Fig.~\ref{fig:reaching}(b), both control laws push the initial bend towards the target until the arm eventually stabilizes in a configuration pointing towards the target. 

%the arm is initialized with an equilibrium configuration under certain voltage boundary values according to~\eqref{eq:voltage_boundary_conditions}. As the arm starts moving, the voltage boundary conditions are set free according to~\eqref{eq:voltage_boundary_conditions_with control}. 
Additionally, we illustrate two properties of the arm: %during the movement among the two control results and that of the benchmark control \eqref{eq:benchmark}: 
(i) tip curvature $\Ktip (t)$ which is defined to be the average curvature of the final $10\%$ of the arm $\Ktip (t) := \tfrac{1}{0.1L}\int_{0.9L}^{L} \kappa (s,t) \ud s$,
% of the segment with length $0.1L$ near the tip as follows:
%\begin{equation*}
%	\Ktip (t) := \frac{1}{0.1L}\int_{0.9L}^{L} \kappa (s,t) \ud s
%\end{equation*}
and (ii) $\bar{s} (t)/L$. %The definition of the point $\bar{s}$ is given in Appendix~\ref{appdx:sensor_model} which describes the location of the arm that is closest to the target. 
The $\bar{s}$ point can be approximately regarded as the position of the propagating bend during the reaching movement.
We see from the trajectories of $\Ktip$ in Fig.~\ref{fig:reaching}(b) that the initial tip curling is released, qualitatively matching bend propagation recordings from biological experiments (Fig.~\ref{fig:reaching}(a)). %The vanishing curling is also reflected in the trajectories of the average tip curvature, as $\Ktip$ goes to zero over time. 
Besides $\Ktip$, the trajectories of $\bar{s}/L$ also show that the reference tracking control successfully tracks the reference.

\section{Conclusion} \label{sec:conclusion}

In this paper, a model for the octopus peripheral nervous system is proposed and integrated within a muscular soft arm. For the proposed neuromuscular control architecture, several analytical results are provided including the characterization of the equilibrium at rest states, and its analytical expression. Numerical results show a qualitative match with experimental observations of curled octopus arms at rest. Besides the rest state, active arm motions are also computationally demonstrated for two proposed control laws. The stability of the closed-loop system under the proposed control laws is discussed. The performance of the two control laws are numerically compared against a benchmark control previously derived. Future work includes investigating possible wave solutions to the neural dynamics, and studying whether mutual inhibition among muscles exist and how it affects the behavior of the arm.

\begin{ack}
The authors are thankful to Dr. Rhanor Gillette’s lab at the University of Illinois and Dr. William Gilly's lab at the Hopkins Marine Station, Stanford University, where the octopus experiments were performed.
\end{ack}

\bibliography{reference}

\appendix

\section{Sketch of proof of proposition~\ref{prop:backstepping}} \label{appdx:backstepping_proof}
We first separate the time-scales of the muscular arm dynamics~\ref{eq:dynamics} and the neural dynamics~\ref{eq:cable_eq_full} by noting that the neural system is `fast' because of the small time constants $\tau, \tilde{\tau}$ (also verified numerically). The `slow' mechanical system can be regarded as frozen with respect to the fast electrical system. Then, the Proposition~\ref{prop:backstepping} is proved by using the method of singular perturbation, as developed in~\citep[Chap. 11]{khalil2002nonlinear}.

Here we give a sketch of the proof by providing various necessary steps. Also, we remove the superscripts $\muscle$ for notational ease. Furthermore, we assume the reference muscle control is a constant in time, i.e. $\bar{u}(s)=c(s)\bar{v}(s)$. Denote the resulting reference voltage as $\Vref=\siginv(\bar{v})$. Then we have the following result.

\begin{lemma}
Consider the neural dynamics~\eqref{eq:cable_eq_full} with the current control~\eqref{eq:reference_tracking_control}, where the reference is constant in time. Then the closed loop neural dynamics is exponentially stable at the equilibrium $(\Vref, bg(\Vref))$. 
\end{lemma}

\textit{Proof:}  We propose the following Lyapunov candidate for the closed loop neural system%~\eqref{eq:cable_eq_full}\eqref{eq:reference_tracking_control}:
\begin{equation*}
\small
	\LyapunovV (V, \Vadapt) = \frac{1}{2}\int_0^L\ \tau\left(V-\Vref\right)^2 + \tauAdapt\left(\Vadapt-bg(\Vref)\right)^2 \ud s
\end{equation*}
It can be readily derived that
\begin{equation}
\scriptsize
	\begin{aligned}
		\frac{\ud \LyapunovV}{\ud t} &= \int_0^L\ -\beta\left(V-\Vref\right)^2 - \left(\Vadapt-bg(\Vref)\right)^2 
		 - \left(V-\Vref\right)\left(\Vadapt-bg(\Vref)\right) \\
		&\qquad + b\left(V-\Vref+\Vadapt-bg(\Vref)\right)\left(g(V)-g(\Vref)\right) \\
		&\qquad +\lambda^2\left(V-\Vref\right)\left(V_{ss} - \Vref_{ss}\right) \ud s \\
		&\leq \int_0^L - (\beta-b)\left|V-\Vref\right|^2 - \left|\Vadapt-bg(\Vref)\right|^2 \\
		&\qquad +(b+1)\left|V-\Vref\right|\left|\Vadapt-bg(\Vref)\right|  - \left|V_s - \Vref_s\right|^2 \ud s\\
		%&\qquad %\lambda^2\left(V-\Vref\right)\left(V_s - \Vref_s\right)\Big|_{s=0}^L 
	\end{aligned} \nonumber
\end{equation}
%Assume the reference voltage $\Vref$ satisfies the same free-free boundary condition as the voltage $V$. Then the second last term is zero.
If $\beta$ satisfies $\beta > \frac{(b+1)^2}{4}+b$, we have $\frac{\ud \LyapunovV}{\ud t}\leq0$ and is only equal to zero at the equilibrium $\left(\Vref, bg(\Vref)\right)$. Therefore, the equilibrium is (locally) asymptotically stable.

Furthermore, for a constant $k=\frac{1}{\tauAdapt}$, if $\beta > \frac{(b+1)^2}{2}+b+\frac{k\tau}{2}$, we have
\begin{equation}
	\frac{\ud \LyapunovV}{\ud t} \leq -k\LyapunovV \nonumber
\end{equation}
and thus the equilibrium is proved to be exponentially stable.

Next, we turn to the mechanical system. It has been shown in~\citep{chang2021controlling} that if the internal muscle couple $\bar{u}$ is a constant, then it is expressible as gradients of an energy function called \textit{muscle stored energy function}). As a consequence, the system maintains its Hamiltonian structure (with damping) and (locally) asymptotic convergence to an equilibrium can be readily shown. The closed-loop Hamiltonian can be taken as a Lyapunov functional, here denoted as $\LyapunovR$. From \citep[Proposition 4.1]{chang2021controlling}, we have that $\frac{\ud \LyapunovR}{\ud t} \leq -\zeta \int_0^L \abs{\frac{1}{\varrho I}p}^2 \dif s$ where $p=\varrho I \theta_t$ is the angular momentum. 

%It is proved in~\cite{chang2021controlling} that $\frac{\ud \LyapunovR}{\ud t}\leq -\zeta|\frac{1}{\varrho I}p|^2$ where $p=\varrho I \theta_t$ is the angular momentum. 

%Let us only consider the case where the muscle couple $u$ described above is a constant. Suppose such $u$ is given as a reference. Denote the resulting reference voltage as $\Vref=\siginv(u)$.

Now, let us consider the following Lyapunov candidate for the coupled system~(\ref{eq:dynamics}, \ref{eq:cable_eq_full}, \ref{eq:neuromuscular_mapping}, \ref{eq:reference_tracking_control}) as $\LyapunovTotal = \LyapunovV + \LyapunovR $.
% given the constant muscle couple $u$:
%\begin{equation*}
%	\LyapunovTotal = \LyapunovR + \LyapunovV
%\end{equation*}
Then along the trajectories of the coupled system, we have
\begin{equation*}
	\begin{aligned}
		\frac{\ud \LyapunovTotal}{\ud t} =& \frac{\ud \LyapunovV}{\ud t} + \int_0^L  \frac{1}{\varrho I}p \left( - \frac{\zeta}{\varrho I} p + \left(\sigma(V)-u\right)_s \right) \ud s
	\end{aligned}
\end{equation*}
The third term is the extra term due to the coupling of the two subsystems. Suppose the model parameters are such that the following inequality holds
\begin{equation}
	\left|\int_0^L \frac{1}{\varrho I}p \left(\sigma(V)-u\right)_s \ud s \right| \leq \beta_0 \int_0^L \frac{1}{\varrho I}\left|p\right|\left|V-\Vref\right| \ud s \nonumber
\end{equation}
for some constant $\beta_0>0$.
Then, we have
\begin{equation}
	\begin{aligned}
	\frac{\ud \LyapunovTotal}{\ud t} \leq&  \int_0^L -\zeta|\frac{1}{\varrho I}p|^2 - \frac{k\tau}{2}\left|V-\Vref\right|^2 + \beta_0\left|\frac{1}{\varrho I}p\right|\left|V-\Vref\right| \\ & -\frac{k\tauAdapt}{2}\left|\Vadapt-bg(\Vref)\right|^2 - \left|V_s-\Vref_s\right|^2  ~\ud s
	\end{aligned} \nonumber
\end{equation}
Thus we finally obtain $\frac{\ud \LyapunovTotal}{\ud t}\leq 0$ for $\beta_0^2 \leq 2\zeta k\tau$. Furthermore, $\frac{\ud \LyapunovTotal}{\ud t} = 0$ only at the equilibrium of the coupled system. Hence, we proved that the coupled system is also (locally) asymptotically stable at the equilibrium.

\end{document}

%% file: definitions.tex
\def\R{{\mathds{R}}}

\def\0{{\mathbb{0}}}
\def\1{{\mathds{1}}}
\newcommand{\abs}[1]{\left| #1 \right|}

\definecolor{db}{RGB}{23,20,119}
\definecolor{dg}{RGB}{2,101,15}

\newtheorem{lemma}{Lemma}[section]
\newtheorem{proposition}{Proposition}[section]

\newtheorem{remark}{Remark}

\usepackage{upgreek}
\newcommand{\dif}{\mathrm{d}}

\newcommand{\set}[1]{\left\{#1\right\}}

\newcommand{\material}[1]{
	\ifthenelse{\equal{#1}{\kappa}}{\upkappa}{
		\ifthenelse{\equal{#1}{\nu}}{\upnu}{
			\ifthenelse{\equal{#1}{\omega}}{\upomega}{
				\ifthenelse{\equal{#1}{\sigma}}{\upsigma}{
					\ifthenelse{\equal{#1}{\theta}}{\uptheta}{
						\mathsf{#1}}}}}}
}

\newcommand{\muscle}{\text{m}}

\newcommand{\ud}{\,\mathrm{d}}

\DeclareMathOperator\arctanh{arctanh}

\newcommand{\rodtip}{\gamma^\text{tip}}
\newcommand{\rodbase}{\gamma^\text{base}}

\newcommand{\Hamiltonian}{\mathcal{H}}

\newcommand{\Vadapt}{W}
\newcommand{\tauAdapt}{\tilde{\tau}}
\newcommand{\LMt}{\mathsf{t}}
\newcommand{\LMb}{\mathsf{b}}
\newcommand{\Vtop}{V^{\LMt}}
\newcommand{\Vbot}{V^{\LMb}}
\newcommand{\Vmuscle}{V^{\muscle}}
\newcommand{\Itop}{I^{\LMt}}
\newcommand{\Ibot}{I^{\LMb}}
\newcommand{\Imuscle}{I^{\muscle}}

\newcommand{\VmA}{W^{\muscle}}
\newcommand{\lambdaA}{\tilde{\lambda}}
\newcommand{\Lyapunov}{\mathcal{E}}
\newcommand{\Ktip}{{\kappa^\text{tip}}}
\newcommand{\LyapunovR}{\Hamiltonian}
\newcommand{\LyapunovV}{\Lyapunov_0}
\newcommand{\LyapunovTotal}{\Lyapunov}
\newcommand{\siginv}{\sigma^{-1}}
\newcommand{\Vref}{\bar{V}}

%% file: neuromuscular_arxiv.bbl
\begin{thebibliography}{53}
\providecommand{\natexlab}[1]{#1}
\providecommand{\url}[1]{\texttt{#1}}
\providecommand{\urlprefix}{URL }
\expandafter\ifx\csname urlstyle\endcsname\relax
  \providecommand{\doi}[1]{doi:\discretionary{}{}{}#1}\else
  \providecommand{\doi}{doi:\discretionary{}{}{}\begingroup
  \urlstyle{rm}\Url}\fi

\bibitem[{Antman(1995)}]{antman1995nonlinear}
Antman, S.S. (1995).
\newblock \emph{Nonlinear Problems of Elasticity}.
\newblock Springer.

\bibitem[{Audu and Davy(1985)}]{audu1985influence}
Audu, M. and Davy, D. (1985).
\newblock The influence of muscle model complexity in musculoskeletal motion
  modeling.

\bibitem[{Aydin et~al.(2019)Aydin, Zhang, Nuethong, Pagan-Diaz, Bashir,
  Gazzola, and Saif}]{aydin2019neuromuscular}
Aydin, O., Zhang, X., Nuethong, S., Pagan-Diaz, G.J., Bashir, R., Gazzola, M.,
  and Saif, M.T.A. (2019).
\newblock Neuromuscular actuation of biohybrid motile bots.
\newblock \emph{Proceedings of the National Academy of Sciences}, 116(40),
  19841--19847.

\bibitem[{Cacace et~al.(2019)Cacace, Lai, and Loreti}]{cacace2019control}
Cacace, S., Lai, A.C., and Loreti, P. (2019).
\newblock Control strategies for an octopus-like soft manipulator.
\newblock In \emph{ICINCO (1)}, 82--90.

\bibitem[{Chang et~al.(2021)Chang, Halder, Gribkova, Tekinalp, Naughton,
  Gazzola, and Mehta}]{chang2021controlling}
Chang, H.S., Halder, U., Gribkova, E., Tekinalp, A., Naughton, N., Gazzola, M.,
  and Mehta, P.G. (2021).
\newblock Controlling a cyberoctopus soft arm with muscle-like actuation.
\newblock In \emph{2021 60th IEEE Conference on Decision and Control (CDC)},
  1383--1390. IEEE.

\bibitem[{Chang et~al.(2022)Chang, Halder, Shih, Naughton, Gazzola, and
  Mehta}]{chang2022energy}
Chang, H.S., Halder, U., Shih, C.H., Naughton, N., Gazzola, M., and Mehta, P.G.
  (2022).
\newblock Energy shaping control of a muscular octopus arm moving in three
  dimensions.
\newblock \emph{arXiv preprint arXiv:2209.04089}.

\bibitem[{Chang et~al.(2020)Chang, Halder, Shih, Tekinalp, Parthasarathy,
  Gribkova, Chowdhary, Gillette, Gazzola, and Mehta}]{chang2020energy}
Chang, H.S., Halder, U., Shih, C.H., Tekinalp, A., Parthasarathy, T., Gribkova,
  E., Chowdhary, G., Gillette, R., Gazzola, M., and Mehta, P.G. (2020).
\newblock Energy shaping control of a cyberoctopus soft arm.
\newblock In \emph{2020 59th IEEE Conference on Decision and Control (CDC)},
  3913--3920. IEEE.

\bibitem[{Di~Clemente et~al.(2021)Di~Clemente, Maiole, Bornia, and
  Zullo}]{di2021beyond}
Di~Clemente, A., Maiole, F., Bornia, I., and Zullo, L. (2021).
\newblock Beyond muscles: role of intramuscular connective tissue elasticity
  and passive stiffness in octopus arm muscle function.
\newblock \emph{Journal of Experimental Biology}, 224(22), jeb242644.

\bibitem[{Ekeberg(1993)}]{ekeberg1993combined}
Ekeberg, {\"O}. (1993).
\newblock A combined neuronal and mechanical model of fish swimming.
\newblock \emph{Biological cybernetics}, 69(5), 363--374.

\bibitem[{FitzHugh(1961)}]{fitzhugh1961impulses}
FitzHugh, R. (1961).
\newblock Impulses and physiological states in theoretical models of nerve
  membrane.
\newblock \emph{Biophysical journal}, 1(6), 445--466.

\bibitem[{Folgheraiter et~al.(2019)Folgheraiter, Keldibek, Aubakir, Gini,
  Franchi, and Bana}]{folgheraiter2019neuromorphic}
Folgheraiter, M., Keldibek, A., Aubakir, B., Gini, G., Franchi, A.M., and Bana,
  M. (2019).
\newblock A neuromorphic control architecture for a biped robot.
\newblock \emph{Robotics and Autonomous Systems}, 120, 103244.

\bibitem[{Gazzola et~al.(2018)Gazzola, Dudte, McCormick, and
  Mahadevan}]{gazzola2018forward}
Gazzola, M., Dudte, L., McCormick, A., and Mahadevan, L. (2018).
\newblock Forward and inverse problems in the mechanics of soft filaments.
\newblock \emph{Royal Society Open Science}, 5(6), 171628.

\bibitem[{Grasso(2014)}]{grasso2014octopus}
Grasso, F.W. (2014).
\newblock The octopus with two brains: how are distributed and central
  representations integrated in the octopus central nervous system.
\newblock \emph{Cephalopod cognition}, 94--122.

\bibitem[{Graziadei and Gagne(1976)}]{graziadei1976sensory}
Graziadei, P. and Gagne, H. (1976).
\newblock Sensory innervation in the rim of the octopus sucker.
\newblock \emph{Journal of morphology}, 150(3), 639--679.

\bibitem[{Gutfreund et~al.(1998)Gutfreund, Flash, Fiorito, and
  Hochner}]{gutfreund1998patterns}
Gutfreund, Y., Flash, T., Fiorito, G., and Hochner, B. (1998).
\newblock Patterns of arm muscle activation involved in octopus reaching
  movements.
\newblock \emph{Journal of Neuroscience}, 18(15), 5976--5987.

\bibitem[{Gutnick et~al.(2020)Gutnick, Zullo, Hochner, and
  Kuba}]{gutnick2020use}
Gutnick, T., Zullo, L., Hochner, B., and Kuba, M.J. (2020).
\newblock Use of peripheral sensory information for central nervous control of
  arm movement by octopus vulgaris.
\newblock \emph{Current Biology}, 30(21), 4322--4327.

\bibitem[{Hanassy et~al.(2015)Hanassy, Botvinnik, Flash, and
  Hochner}]{hanassy2015stereotypical}
Hanassy, S., Botvinnik, A., Flash, T., and Hochner, B. (2015).
\newblock Stereotypical reaching movements of the octopus involve both bend
  propagation and arm elongation.
\newblock \emph{Bioinspiration \& biomimetics}, 10(3), 035001.

\bibitem[{Hatze(1977)}]{hatze1977myocybernetic}
Hatze, H. (1977).
\newblock A myocybernetic control model of skeletal muscle.
\newblock \emph{Biological cybernetics}, 25(2), 103--119.

\bibitem[{Hodgkin and Huxley(1952)}]{hodgkin1952quantitative}
Hodgkin, A.L. and Huxley, A.F. (1952).
\newblock A quantitative description of membrane current and its application to
  conduction and excitation in nerve.
\newblock \emph{The Journal of physiology}, 117(4), 500.

\bibitem[{Ijspeert et~al.(2007)Ijspeert, Crespi, Ryczko, and
  Cabelguen}]{ijspeert2007swimming}
Ijspeert, A.J., Crespi, A., Ryczko, D., and Cabelguen, J.M. (2007).
\newblock From swimming to walking with a salamander robot driven by a spinal
  cord model.
\newblock \emph{science}, 315(5817), 1416--1420.

\bibitem[{Izhikevich(2007)}]{izhikevich2007dynamical}
Izhikevich, E.M. (2007).
\newblock \emph{Dynamical systems in neuroscience}.
\newblock MIT press.

\bibitem[{Kennedy et~al.(2020)Kennedy, Buresch, Boinapally, and
  Hanlon}]{kennedy2020octopus}
Kennedy, E.L., Buresch, K.C., Boinapally, P., and Hanlon, R.T. (2020).
\newblock Octopus arms exhibit exceptional flexibility.
\newblock \emph{Scientific reports}, 10(1), 1--10.

\bibitem[{Khalil(2002)}]{khalil2002nonlinear}
Khalil, H.K. (2002).
\newblock \emph{Nonlinear systems third edition}.
\newblock Patience Hall.

\bibitem[{Kier(2016)}]{kier2016musculature}
Kier, W.M. (2016).
\newblock The musculature of coleoid cephalopod arms and tentacles.
\newblock \emph{Frontiers in cell and developmental biology}, 4, 10.

\bibitem[{Kier and Stella(2007)}]{kier2007arrangement}
Kier, W.M. and Stella, M.P. (2007).
\newblock The arrangement and function of octopus arm musculature and
  connective tissue.
\newblock \emph{Journal of morphology}, 268(10), 831--843.

\bibitem[{Koch(1984)}]{koch1984cable}
Koch, C. (1984).
\newblock Cable theory in neurons with active, linearized membranes.
\newblock \emph{Biological cybernetics}, 50(1), 15--33.

\bibitem[{Levy et~al.(2017)Levy, Nesher, Zullo, and Hochner}]{levy2017motor}
Levy, G., Nesher, N., Zullo, L., and Hochner, B. (2017).
\newblock Motor control in soft-bodied animals: the octopus.
\newblock In \emph{The Oxford Handbook of Invertebrate Neurobiology}.

\bibitem[{Liu et~al.(2008)Liu, Habib, Watanabe, and Izumi}]{liu2008central}
Liu, G.L., Habib, M.K., Watanabe, K., and Izumi, K. (2008).
\newblock Central pattern generators based on matsuoka oscillators for the
  locomotion of biped robots.
\newblock \emph{Artificial Life and Robotics}, 12(1), 264--269.

\bibitem[{Mather(2021)}]{mather2021octopus}
Mather, J. (2021).
\newblock Octopus consciousness: the role of perceptual richness.
\newblock \emph{NeuroSci}, 2(3), 276--290.

\bibitem[{Mather(1998)}]{mather1998octopuses}
Mather, J.A. (1998).
\newblock How do octopuses use their arms?
\newblock \emph{Journal of Comparative Psychology}, 112(3), 306.

\bibitem[{Matsuoka(1984)}]{matsuoka1984dynamic}
Matsuoka, K. (1984).
\newblock The dynamic model of binocular rivalry.
\newblock \emph{Biological cybernetics}, 49(3), 201--208.

\bibitem[{Matzner et~al.(2000)Matzner, Gutfreund, and
  Hochner}]{matzner2000neuromuscular}
Matzner, H., Gutfreund, Y., and Hochner, B. (2000).
\newblock Neuromuscular system of the flexible arm of the octopus:
  physiological characterization.
\newblock \emph{Journal of neurophysiology}, 83(3), 1315--1328.

\bibitem[{Nagumo et~al.(1962)Nagumo, Arimoto, and Yoshizawa}]{nagumo1962active}
Nagumo, J., Arimoto, S., and Yoshizawa, S. (1962).
\newblock An active pulse transmission line simulating nerve axon.
\newblock \emph{Proceedings of the IRE}, 50(10), 2061--2070.

\bibitem[{Nesher et~al.(2020)Nesher, Levy, Zullo, and
  Hochner}]{nesher2020octopus}
Nesher, N., Levy, G., Zullo, L., and Hochner, B. (2020).
\newblock Octopus motor control.
\newblock \emph{Oxford Research Encyclopedia of Neuroscience. Oxford University
  Press, USA}.

\bibitem[{Nesher et~al.(2019)Nesher, Maiole, Shomrat, Hochner, and
  Zullo}]{nesher2019synaptic}
Nesher, N., Maiole, F., Shomrat, T., Hochner, B., and Zullo, L. (2019).
\newblock From synaptic input to muscle contraction: arm muscle cells of
  octopus vulgaris show unique neuromuscular junction and
  excitation--contraction coupling properties.
\newblock \emph{Proceedings of the Royal Society B}, 286(1909), 20191278.

\bibitem[{Packard and Sanders(1971)}]{packard1971body}
Packard, A. and Sanders, G.D. (1971).
\newblock Body patterns of octopus vulgaris and maturation of the response to
  disturbance.
\newblock \emph{Animal Behaviour}, 19(4), 780--790.

\bibitem[{Polykretis et~al.(2022)Polykretis, Supic, and
  Danielescu}]{polykretis2022bioinspired}
Polykretis, I., Supic, L., and Danielescu, A. (2022).
\newblock Bioinspired smooth neuromorphic control for robotic arms.
\newblock \emph{arXiv preprint arXiv:2209.02787}.

\bibitem[{Rall(1962)}]{rall1962theory}
Rall, W. (1962).
\newblock Theory of physiological properties of dendrites.
\newblock \emph{Annals of the New York Academy of Sciences}, 96(4), 1071--1092.

\bibitem[{Rowell(1963)}]{rowell1963excitatory}
Rowell, C.F. (1963).
\newblock Excitatory and inhibitory pathways in the arm of octopus.
\newblock \emph{Journal of Experimental Biology}, 40(2), 257--270.

\bibitem[{Sfakiotakis and Tsakiris(2007)}]{sfakiotakis2007neuromuscular}
Sfakiotakis, M. and Tsakiris, D.P. (2007).
\newblock Neuromuscular control of reactive behaviors for undulatory robots.
\newblock \emph{Neurocomputing}, 70(10-12), 1907--1913.

\bibitem[{Sumbre et~al.(2001)Sumbre, Gutfreund, Fiorito, Flash, and
  Hochner}]{sumbre2001control}
Sumbre, G., Gutfreund, Y., Fiorito, G., Flash, T., and Hochner, B. (2001).
\newblock Control of octopus arm extension by a peripheral motor program.
\newblock \emph{Science}, 293(5536), 1845--1848.

\bibitem[{Tian and Lu(2015)}]{tian2015simulation}
Tian, J. and Lu, Q. (2015).
\newblock Simulation of octopus arm based on coupled cpgs.
\newblock \emph{Journal of Robotics}, 2015.

\bibitem[{Tuckwell(1988)}]{tuckwell1988introduction}
Tuckwell, H.C. (1988).
\newblock \emph{Introduction to theoretical neurobiology}.
\newblock New York: Cambridge University Press.

\bibitem[{Wang et~al.(2021)Wang, Halder, Chang, Gazzola, and
  Mehta}]{wang2021optimal}
Wang, T., Halder, U., Chang, H.S., Gazzola, M., and Mehta, P.G. (2021).
\newblock Optimal control of a soft cyberoctopus arm.
\newblock In \emph{2021 American Control Conference (ACC)}, 4757--4764. IEEE.

\bibitem[{Wang et~al.(2022{\natexlab{a}})Wang, Halder, Gribkova, Gazzola, and
  Mehta}]{wang2022control}
Wang, T., Halder, U., Gribkova, E., Gazzola, M., and Mehta, P.G.
  (2022{\natexlab{a}}).
\newblock Control-oriented modeling of bend propagation in an octopus arm.
\newblock In \emph{American Control Conference (ACC)}, 1359--1366. IEEE.

\bibitem[{Wang et~al.(2022{\natexlab{b}})Wang, Halder, Gribkova, Gillette,
  Gazzola, and Mehta}]{wang2022sensory}
Wang, T., Halder, U., Gribkova, E., Gillette, R., Gazzola, M., and Mehta, P.G.
  (2022{\natexlab{b}}).
\newblock A sensory feedback control law for octopus arm movements.
\newblock In \emph{2022 61th IEEE Conference on Decision and Control (CDC)
  (accepted)}. IEEE.

\bibitem[{Wang et~al.(2020)Wang, Taghvaei, and Mehta}]{wang2020bio}
Wang, T., Taghvaei, A., and Mehta, P.G. (2020).
\newblock Bio-inspired learning of sensorimotor control for locomotion.
\newblock In \emph{2020 American Control Conference (ACC)}, 2188--2193. IEEE.

\bibitem[{Wells(1978)}]{wells1978octopus}
Wells, M.J. (1978).
\newblock \emph{Octopus: physiology and behaviour of an advanced invertebrate}.
\newblock Springer.

\bibitem[{Yekutieli et~al.(2005{\natexlab{a}})Yekutieli, Sagiv-Zohar, Aharonov,
  Engel, Hochner, and Flash}]{yekutieli2005dynamic}
Yekutieli, Y., Sagiv-Zohar, R., Aharonov, R., Engel, Y., Hochner, B., and
  Flash, T. (2005{\natexlab{a}}).
\newblock Dynamic model of the octopus arm. i. biomechanics of the octopus
  reaching movement.
\newblock \emph{Journal of neurophysiology}, 94(2), 1443--1458.

\bibitem[{Yekutieli et~al.(2005{\natexlab{b}})Yekutieli, Sagiv-Zohar, Hochner,
  and Flash}]{yekutieli2005dynamic2}
Yekutieli, Y., Sagiv-Zohar, R., Hochner, B., and Flash, T.
  (2005{\natexlab{b}}).
\newblock Dynamic model of the octopus arm. ii. control of reaching movements.
\newblock \emph{Journal of neurophysiology}, 94(2), 1459--1468.

\bibitem[{Yoram et~al.(2002)Yoram, German, Tamar, and Binyamin}]{yoram2002move}
Yoram, Y., German, S., Tamar, F., and Binyamin, H. (2002).
\newblock How to move with no rigid skeleton? the octopus has the answers.
\newblock \emph{Biologist}, 49(6), 250--254.

\bibitem[{Young(1971)}]{young1971anatomy}
Young, J.Z. (1971).
\newblock \emph{The anatomy of the nervous system of Octopus vulgaris}.
\newblock Oxford: Clarendon Press.

\bibitem[{Zhang et~al.(2019)Zhang, Chan, Parthasarathy, and
  Gazzola}]{zhang2019modeling}
Zhang, X., Chan, F.K., Parthasarathy, T., and Gazzola, M. (2019).
\newblock Modeling and simulation of complex dynamic musculoskeletal
  architectures.
\newblock \emph{Nature Communications}, 10(1), 1--12.

\end{thebibliography}
